\documentclass[osajnl,twocolumn,showpacs,superscriptaddress,10pt]{revtex4-1}
\usepackage{amsmath,amssymb,graphicx}

\newcommand{\sgn}{\mathrm{sgn}}

\begin{document}%\twocolumn[

\title{Passive mode-locking under higher order effects}

\author{Theodoros P. Horikis}\email{Corresponding author: horikis@uoi.gr}
\affiliation{Department of Mathematics, University of Ioannina, Ioannina 45110, Greece}

\author{Mark J. Ablowitz}
\affiliation{Department of Applied Mathematics, University of Colorado, 526
UCB, Boulder, CO 80309-0526, USA}

\begin{abstract}
The response of a passive mode-locking mechanism, where  gain and spectral
filtering are saturated with the energy and loss saturated with the power,
is examined under the presence of higher order effects. These include third
order dispersion, self-steepening and Raman gain.
The locking mechanism is maintained even with these terms; mode-locking
occurs for both the
anomalous and normal regimes. In the anomalous regime, these
perturbations are found to affect the speed but not the structure of the
(locked) pulses. In fact, these pulses behave like solitons of a classical
nonlinear Schr\"odinger equation and as such a soliton perturbation theory is used to verify the numerical observations. In the normal regime, the effect of the perturbations is small, in line with recent
experimental observations. The results in the normal regime are verified mathematically using a WKB type asymptotic theory. Finally, bi-solitons are found to behave as dark
solitons on top of a stable background and are significantly affected  by these
perturbations.
\end{abstract}

%]

\maketitle %% required

\section{Introduction}
Ultrashort pulses are used frequently in modern applications ranging from
femtochemistry and medical imaging to micro-machining and optical
communications. Consequently the study and modeling of such pulses is an
important area of research. Kerr-lens mode locking (KLM) is a common
technique used to generate these pulses. In fact, following the discovery of
KLM mode-locked (ML) lasers have revolutionized the field of ultrafast
science and their performance has led to their widespread
use.

Mode-locked lasers operate under the requirement that sufficient amounts of
gain and loss are present, otherwise, they are subject to radiation-mode
instability. In addition, passive mode-locking generally utilizes fast
saturable absorbers which are effectively simulated by KLM. Saturable
absorbers are commonly modeled by a transfer function and as such loss is
placed periodically during the evolution and not as part of a
distributed model, i.e., an evolution equation. However, the qualitative features are the
same in both cases \cite{horikis2} thus indicating that distributive models
are very good descriptions of modes in mode-locked lasers.

Modeling these lasers must take into account their physical properties. For this, the distributive power-energy saturation (PES) model was recently
introduced \cite{horikis5}. This system is a variant of the well known nonlinear
Schr\"odinger (NLS) equation with additional terms to account for gain and
spectral filtering, saturated with energy, and loss, saturated with power. It
is a generalization of the commonly used master-equation
\cite{master1,master3} which is obtained as a limiting case. Mode-locking is
much more robust in the PES model than in the master-equation. Indeed, while
in the master-equation mode-locking is achieved only for a narrow range of
parameters \cite{kutz,kutz2}, the PES equation has only one key requirement
for mode-locking: sufficient gain. Another important feature of the PES equation
is that it leads to physically relevant results for both the anomalous and
normal dispersive regimes.

Solitons in ML lasers are fundamentally different in the two dispersive
regimes. Pulse formation, in the anomalous regime, is typically dominated by
the interplay between dispersion and nonlinearity. Suitable gain media and an
effective saturable absorber are required for initiation of pulsed operation.
In contrast, pulses found in the normal regime are positively chirped
throughout the cavity \cite{ilday}. They are comprised of an approximately
parabolic temporal amplitude profile near the peak of the pulse with a
transition to steep decay \cite{ilday}. Pulses in the normal regime have a
much larger temporal width than those in the anomalous regime.

Here we study the response of a ML laser under higher order effects using the
PES equation. These effects include third-order dispersion (TOD),
self-steepening and Raman gain. Previous studies, in the anomalous dispersion
regime, include variants of the master-equation \cite{haus,lin} and
generalized Ginzburg-Landau type equations \cite{kalashnikov,song,sorokin}.
In the normal regime recent experiments \cite{oktem} indicate that
similariton (self-similar) evolution in these fiber lasers results in
localized modes, i.e. solitons, which are broad in the time domain; they
 are robust against perturbations. The effect of TOD on
self-similar evolution and parabolic pulses was studied in Refs.
\cite{dudley,bale,zhang}.

Here we focus on the equation
\begin{align}
i\frac{{\partial u}}{{\partial z}} + \frac{{{d_0}}}{2}\frac{{{\partial ^2}u}}{{\partial
{t^2}}} + |u{|^2}u &= \frac{{igu}}{{1 + \epsilon E}} + \frac{{i\tau }}{{1 + \epsilon
E}}\frac{{{\partial ^2}u}}{{\partial {t^2}}}- \frac{{ilu}}{{1 + \delta P}}\nonumber\\ &+
i\beta \frac{{{\partial ^3}u}}{{\partial {t^3}}} -i\gamma \frac{\partial (|u|^2u)}{\partial
t} +
Ru\frac{{\partial
(|u|^2)}}{{\partial t}}
\label{pes}
\end{align}
where $g$ represents gain saturated with the energy $E=\int_{-\infty}^\infty
|u|^2\; dt$, $\tau$ is spectral filtering, $l$ is the gain saturated with
power $P=|u|^2$, $\beta$ represents TOD, $\gamma$ self-steepening and $R$ the
Raman gain. The parameters $\epsilon>0$ and $\delta>0$ correspond to a
measure of the saturation energy and power of the system, respectively.
Unless otherwise stated all parameters, except $d_0$, are positive. When
$\beta=\gamma=R=0$ this equation is the PES model. Finally, other linear and higher order nonlinear dissipative terms, such as terms proportional to $iu$ and
$i|u|^2u$ are included in Eq. \eqref{pes} as limiting cases in the expansion for the saturable terms and will not be considered here.

\section{The anomalous dispersion regime}

\begin{figure}[!htbp]
    \centering
    \includegraphics[width=2.5in]{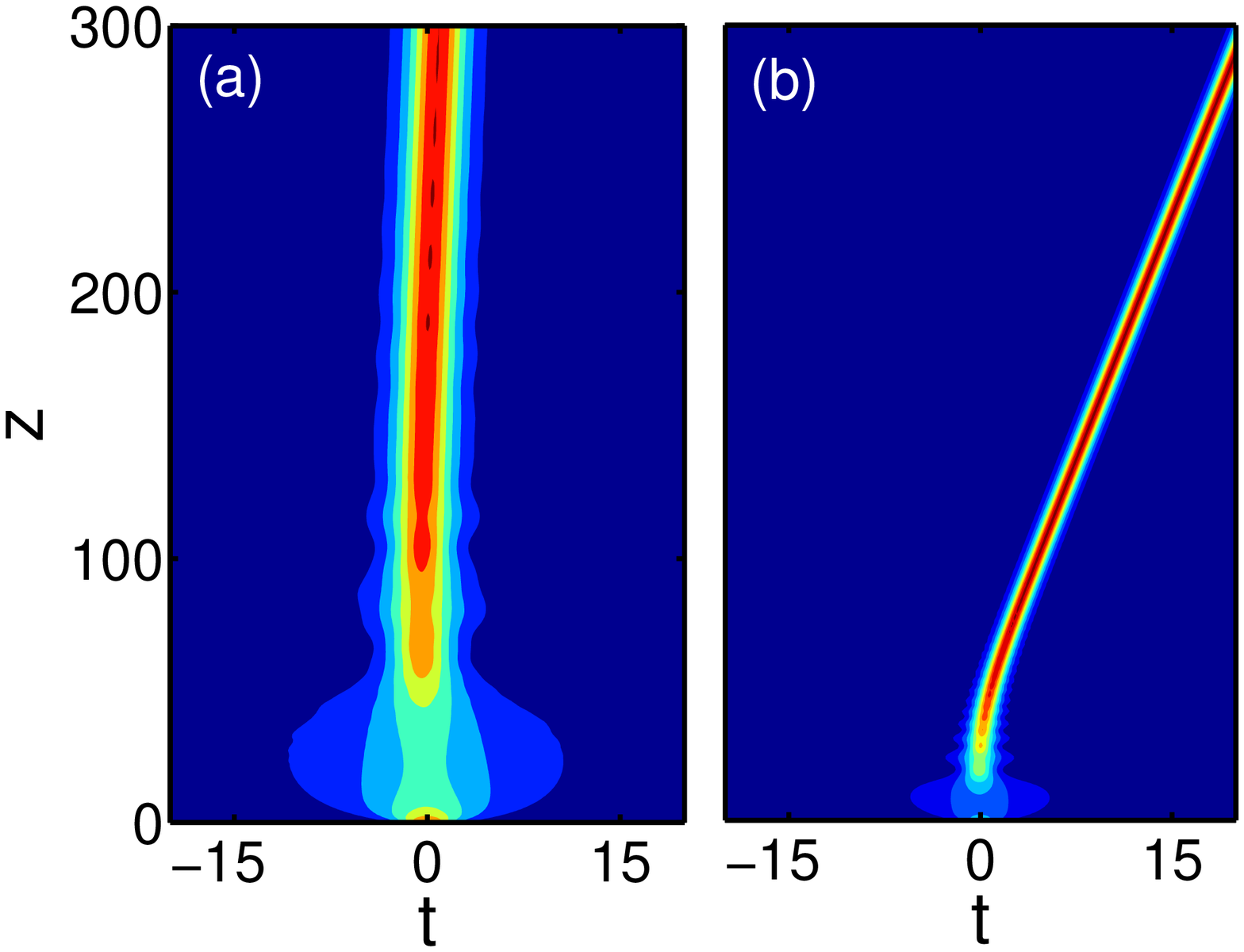}\\[-8pt]
    \includegraphics[width=2.5in]{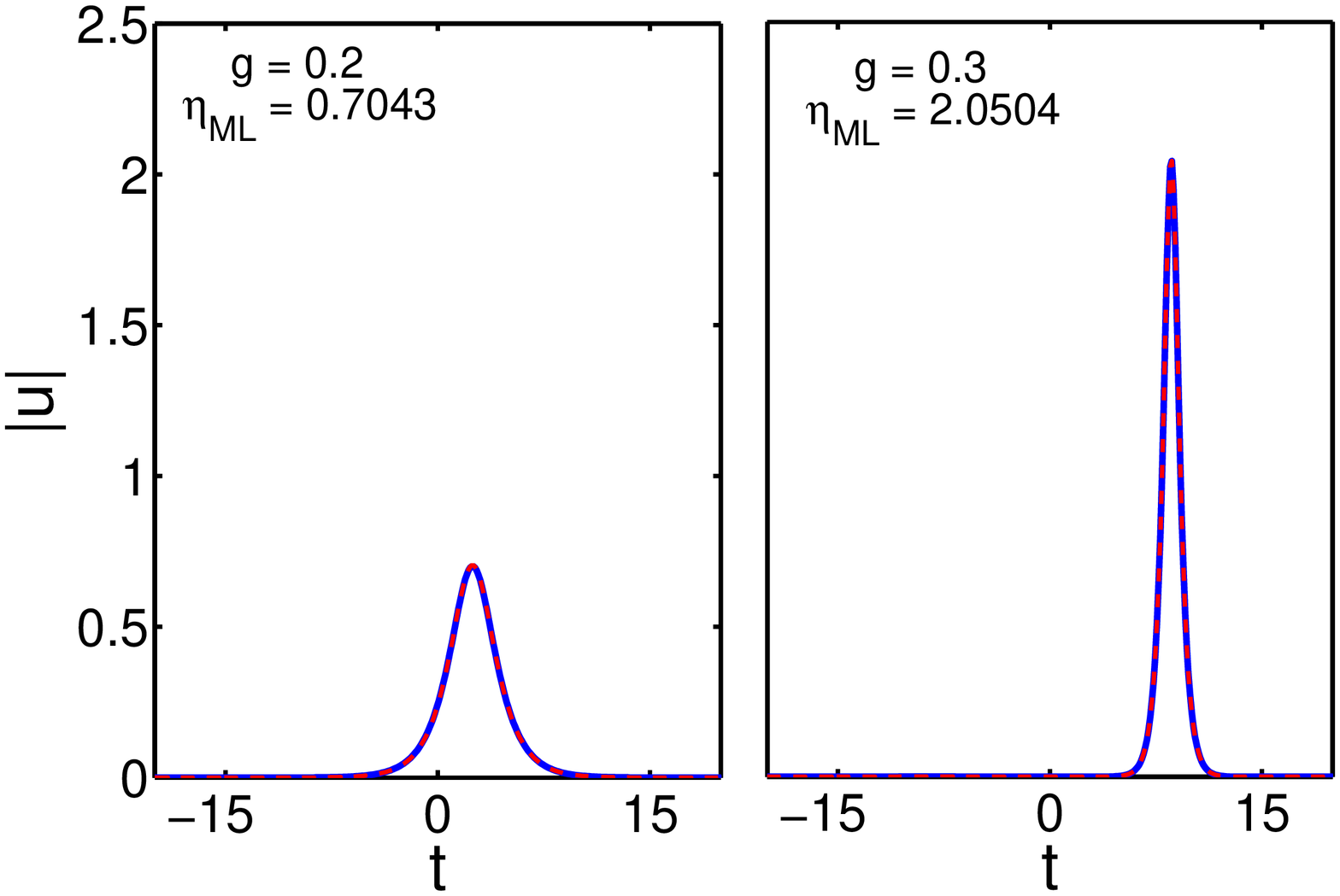}
    \caption{(Color online) Mode-locking under TOD with
     (a) $g=0.2$ and (b) $g=0.3$. The bottom figures show the pulse at a
     distance $z=150$ (a sufficient distance of propagation after mode-locking has
     occurred).
     The dashed lines correspond to the relative classical NLS solitons.}
    \label{tod}
\end{figure}

Here we consider Eq. \eqref{pes} with $d_0>1$ and we start our
analysis by analyzing the effect of TOD. To do so, we fix the
parameters such that $\tau=l=0.1$, $d_0=\epsilon=\delta=1$ and $\beta=0.01$ while
$\gamma=R=0$ and vary the gain parameter $g$. A unit gaussian is evolved
under Eq. \eqref{pes} with the resulting evolution shown in Fig. \ref{tod}.

As seen here, the initial gaussian undergoes a locking procedure and
subsequently  locks to a pulse of constant height/shape. After that happens
the effect of TOD shifts the soliton with constant velocity (seen as a
straight line for the soliton center in the top of Fig. \ref{tod}).
Interestingly the resulting mode is a solution of the unperturbed NLS equation.
Furthermore, when the gain is increased the angle of the displacement becomes
more acute. This is reminiscent of the response of a single soliton of the
classical NLS equation to TOD. Indeed, recall that under TOD the NLS soliton
is slowed down and as a result the soliton peak is shifted by an amount that
increases linearly with distance \cite{book2,book1}.

\begin{figure}[!htbp]
    \centering
    \includegraphics[width=2.5in]{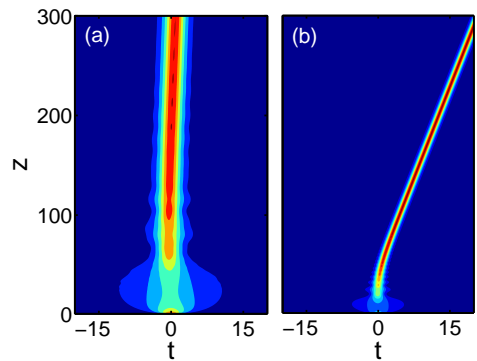}
    \caption{(Color online) Mode-locking under self-steepening with
    (a) $g=0.2$ and (b) $g=0.3$. Profiles
    of mode-locked pulses resemble those of Fig. \ref{tod} (bottom).}
    \label{steepening}
\end{figure}

Similarly, we repeat the analysis for $\beta=R=0$ and $\gamma=0.01$ so as to
study the response to self-steepening. The resulting evolution is depicted in
Fig. \ref{steepening}.

The similarities between Figs. \ref{tod} and \ref{steepening} are remarkable.
Indeed, after mode-locking occurs the pulse is shifted linearly in a
way that is  similar to the way a classical NLS soliton propagates
under this effect.

\begin{figure}[!htbp]
    \centering
    \includegraphics[width=2.5in]{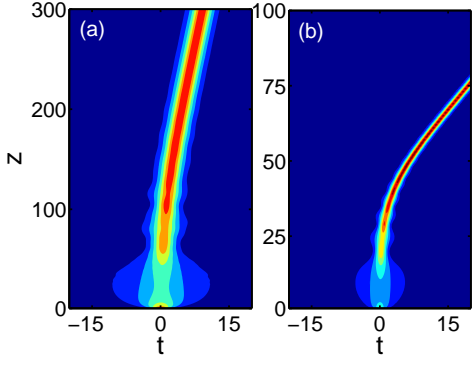}
    \caption{(Color online) Mode-locking under Raman gain with
    (a) $g=0.2$ and (b) $g=0.3$. Profiles
    of mode-locked pulses resemble those of Fig. \ref{tod} (bottom).}
    \label{raman}
\end{figure}

Finally, we repeat the analysis for $\beta=\gamma=0$ and $R=0.01$ so as to
study the response to the Raman gain. The resulting evolution is depicted in
Fig. \ref{raman}.

The phenomenon is repeated only now the displacement is sharper. Again, recall that Raman gain on solitons of the classical NLS equation gives rise to the so-called self-frequency shift \cite{book2,book1}. Note the size of the pulse amplitude is largely unaffected.

It is also important to stress that this locking behavior strongly depends on
the saturation terms. If these terms/effects are absent more complicated
dynamics can occur \cite{sorokin}.

\subsection{Perturbation theory}

\begin{figure}[!htbp]
    \centering
    \includegraphics[width=3in]{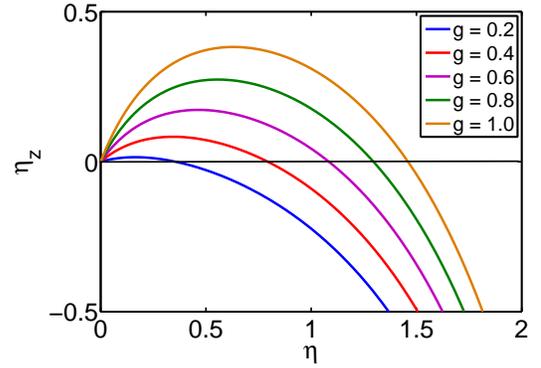}
    \caption{Phase portrait of the amplitude equation for several values of $g$ illustrating the single stable equilibrium. Here $l=0.1$.}
    \label{etaphase}
\end{figure}

Now we shall explain the above features using soliton perturbation theory valid
for small $\beta$, $\gamma$, $R$. This approach leads to very
good qualitative results since when $\beta=\gamma=R=0$ general initial
conditions, such as unit gaussians, will evolve to solitons of
the unperturbed classical NLS equation \cite{horikis7}. To do this, we
consider a solution ansatz for Eq. \eqref{pes} of the form
\[
u(z,t)=\eta(z)\mathrm{sech}[\eta(z)(t-t_0(z))]e^{i\phi(z)-i\sigma(z)t}
\]
Under the effect of the perturbed PES the soliton parameters evolve according
to \cite{horikis1,book1,book2}
\begin{subequations}
\begin{align}
  \frac{{d\eta }}{{dz}} &= \frac{{2g\eta }}{{1 + 2\epsilon\eta }} - \frac{{2\tau ({\eta ^2}
  +
  3{\sigma ^2})}}{{3(1 + 2\epsilon\eta )}}\eta  - \frac{{2l{{\tanh }^{ - 1}}\left(
  {\frac{{\sqrt \delta  \eta }}{{\sqrt {1 + \delta {\eta ^2}} }}} \right)}}{{\sqrt {\delta
  +
  {\delta ^2}{\eta ^2}} }}
  \label{pert1}\\
  \frac{{d\sigma }}{{dz}} &=  - \frac{{4\tau {\eta ^2}\sigma }}{{3(1 + 2\epsilon\eta )}} -
  \frac{8}{{15}}R{\eta ^4} \label{pert2}\\
  \frac{{d{t_0}}}{{dz}} &=  - \sigma  + \beta (\eta ^2+3\sigma^2)+\gamma\eta^2
  \label{pert3}\\
  \frac{{d\phi }}{{dz}} &= \beta \sigma (3{\eta ^2} + {\sigma ^2}) +\gamma\eta^2 +
  \frac{1}{2}({\eta ^2} -
  {\sigma ^2}) + {t_0}\frac{{d\sigma }}{{dz}}
  \label{pert4}
\end{align}
\label{pert}
\end{subequations}
When the pulse has mode-locked, the zeroes of the right hand side of Eq.
\eqref{pert1}-\eqref{pert2} give the
stationary solutions for $\eta$ and $\sigma$. We denote the mode locked value of
$\eta$ and $\sigma$ as $\eta_{ML}$ and $\sigma_{ML}$, respectively.
As such when the equilibrium has been reached
\[
\sigma\equiv\sigma_{ML}=-\frac{2R(1+2\epsilon\eta)\eta^2}{5\tau}
\]
this results in the following equation for $\eta$
\begin{align*}
\frac{d\eta}{dz}= \frac{{2g\eta }}{{1 + 2\epsilon\eta }} &-\frac{2 \left[25\tau ^2 \eta ^2+12 (2 \epsilon R \eta +R)^2\eta ^4\right]}{75 (\tau +2\epsilon\tau \eta  )} \\ &- \frac{{2l{{\tanh }^{ - 1}}\left(
  {\frac{{\sqrt \delta  \eta }}{{\sqrt {1 + \delta {\eta ^2}} }}} \right)}}{{\sqrt {\delta
  +
  {\delta ^2}{\eta ^2}} }}
\end{align*}
Clearly, $\eta$ has at least one equilibrium at $\eta=0$. To understand its stability
consider the first derivative of $\eta_z$ with respect to $\eta$
evaluated at $\eta=0$,
\[
\left.\frac{d}{d\eta} \left(\frac{d\eta}{dz}\right)\right|_{\eta=0} = 2(g-l)
\]
When $g>l, \eta=0$ is an unstable equilibrium while for $g<l, \eta= 0$ is  stable.
The phase portrait for various values of $g<l$  indicates  that $\eta_z$
is a monotonic decreasing function in $\eta$ and so $\eta \rightarrow 0$ as $z
\rightarrow \infty$ for any initial condition $\eta>0$.  Physically this
corresponds to loss overtaking gain and the pulse decays.  For $g>l$,
$\eta_z$ is initially positive and as $\eta$ increases the filtering
term becomes dominant
and $\eta_z$ takes on negative values.  Thus, there must exist another
equilibrium.  Plotting the phase portrait for various values of $g$ for $g>l$ shows
there is a single stable equilibrium for $\eta>0$, as shown in Fig. \ref{etaphase}.

As $\eta$ approaches its stable equilibrium, $\eta_{ML} \neq 0$, and $\sigma$ approaches its stable equilibrium, $\sigma_{ML}$, the system of equations \eqref{pert} tend to
\[
\left. \frac{d\eta}{dz}\right|_{\eta=\eta_{ML}} =
\left.\frac{d\sigma}{dz}\right|_{\sigma=\sigma_{ML}} = 0
\]
which means the solition tends to a particular NLS soliton of speed zero and
height/width determined by $\eta_{ML}$, thus suggesting the mode-locking capabilities
of Eq. (\ref{pes}). In fact, $\eta_{ML}$ is an attractor that any arbitrary initial
condition will eventually converge to, even when it is far way from its soliton
solution \cite{horikis5,horikis7}.

After $\eta = \eta_{ML}$ and mode-locking has been completed,  we
consider Eq. \eqref{pert3}-\eqref{pert4}, which, interestingly, are independent of
$g, \tau, l$.
This now explains the above dynamics. Indeed, when
$\eta=\eta_{ML}$ (and the mode-locking mechanism has been completed) we get
by direct integration of Eq. \eqref{pert3}
\begin{align}
t_0(z) &= t_0(0)+\left\{\left(\beta +\gamma +\frac{2 R}{5 \tau }\right)\eta_{ML} ^2\right.\nonumber\\
&\left.+ \frac{4 R \left[3 \beta  R (2\epsilon \eta_{ML}  +1)^2\eta_{ML}+5\epsilon \tau  \right]\eta_{ML} ^3}{25 \tau ^2}\right\}z
\label{t0}
\end{align}
The equations for the equilibria give now more insight as to the way each term affects the propagation. For instance, in the absence of spectral filtering there still exists a stable (nonzero) equilibrium for $\eta$ but not for $\sigma$. While the amplitude is affected only in its magnitude the other quantities change in a more prominent way. Indeed, by direct integration of Eqs. \eqref{pert2} and \eqref{pert3} we obtain at $\eta=\eta_{ML}$
\begin{align*}
\sigma(z) &= - \frac{8}{{15}}R{\eta_{ML}^4}z+\sigma(0)\\
t_0(z) &= \frac{64}{225} R^2 \beta \eta_{ML}^8 z^3
-\frac{4}{15} R \eta_{ML}^4 [6 \beta \sigma(0)-1] z^2\nonumber\\
&+[(\beta+\gamma)\eta_{ML}^2+3\beta\sigma^2(0)-3\sigma(0)]z+t_0(0)
\end{align*}
which suggests that the trajectory the soliton follows is no longer linear. Similarly when the Raman gain is not considered the only equilibrium is found for $\sigma\equiv0$.  Clearly when $\tau=R=0$ the soliton moves with constant velocity equal to
$(\beta+\gamma)\eta_{ML}^2$ ($\sigma\equiv 0$).
Also, note that as supported by  the numerical
computations the
final amplitude of the pulse is not affected by the parameters $\beta, \gamma$.

\begin{figure}[!htbp]
    \centering
    \includegraphics[width=3in]{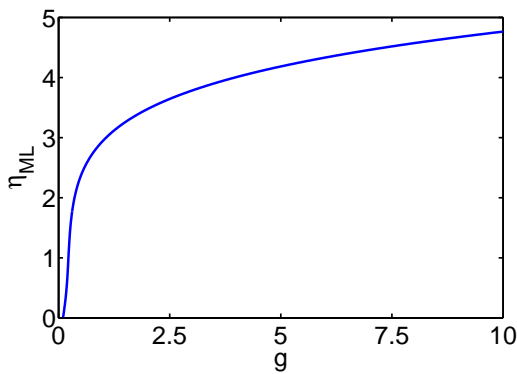}
    \caption{(Color online) The amplitude dependance on
    gain based on Eq. \eqref{pert1}.}
    \label{eta_g}
\end{figure}

The increase of $\eta$ with $g$ also explains the pronounced
difference in the speed of the pulse when $g$ is increased. In Fig.
\ref{eta_g} we depict the dependance of the amplitude to the gain parameter,
which attests to that.
In fact, Eq. \eqref{t0} clearly suggests that the trajectories are highly dependent
on powers of $\eta_{ML}$. Hence, any increase on this amplitude is enhanced
(when $\eta_{ML}>1$ or $g>0.22$) making the response to the perturbations
more pronounced. This analysis also suggests that nonlinear phase shifts
cannot compensate for TOD (or self-steepening) in this regime, unlike in the
normal regime \cite{zhou}.

Thus, the mode-locking mechanism can now be summarized as follows: (i) solve
Eqs. \eqref{pert1}-\eqref{pert2}, to determine $\eta_{ML}$ and $\sigma_{ML}$; (ii)
assume the mode-locking mechanism has been completed;  (iii) treat the pulses
as classical NLS solitons; and
(iv) use Eq. \eqref{t0} and \eqref{pert4} to determine the
adiabatic change of the rest of the soliton parameters.

\section{The normal dispersion regime}

We turn our attention to the normal regime ($d_0=-1$). The main
difference these pulses exhibit from their anomalous regime counterparts is
that they are highly chirped, very wide (almost parabolic) pulses in the time
domain; the modes depend critically on the
values of $g,\tau,l$.

\begin{figure}[!htbp]
    \centering
    \includegraphics[width=2.5in]{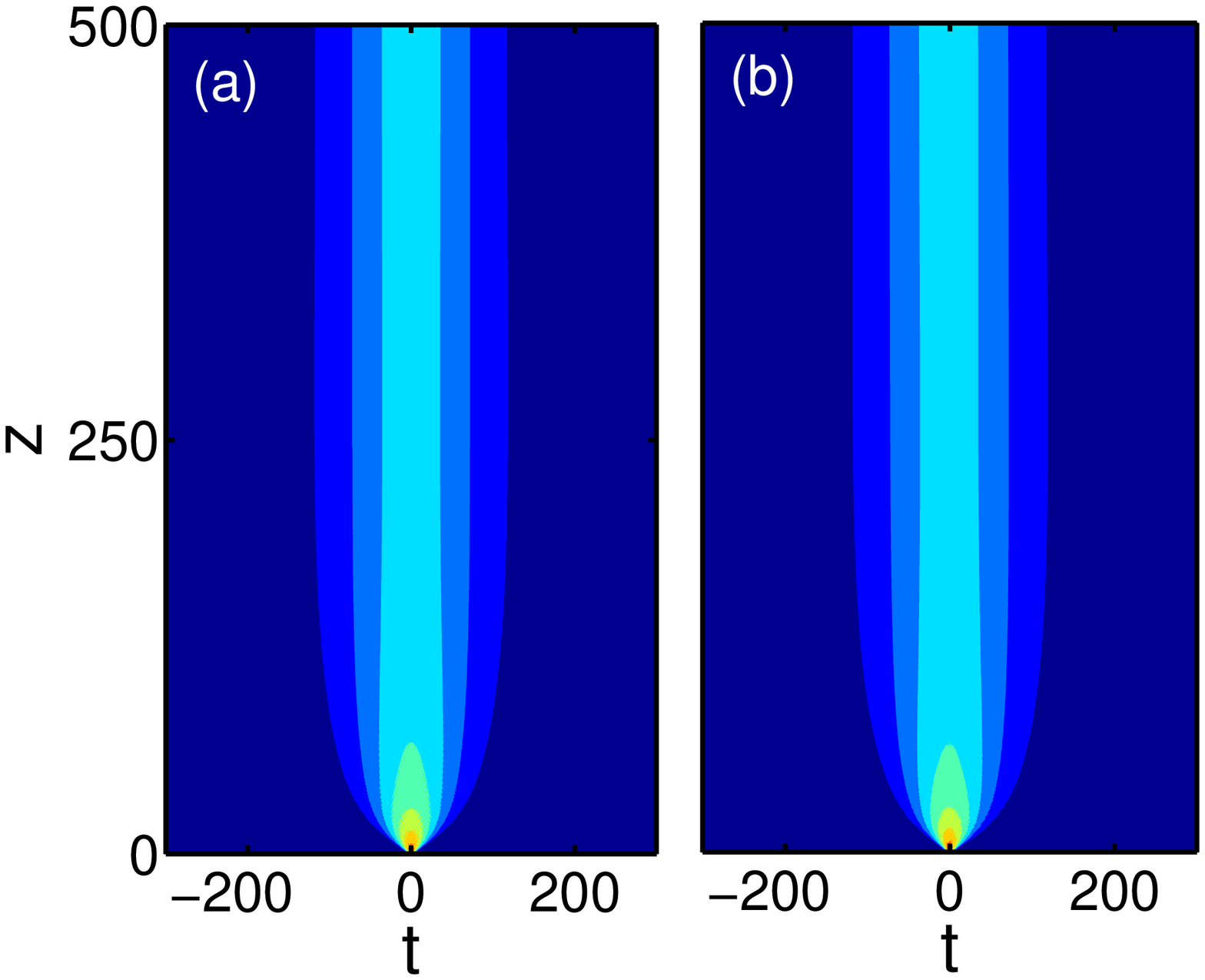}\\[-8pt]
    \includegraphics[width=2.5in]{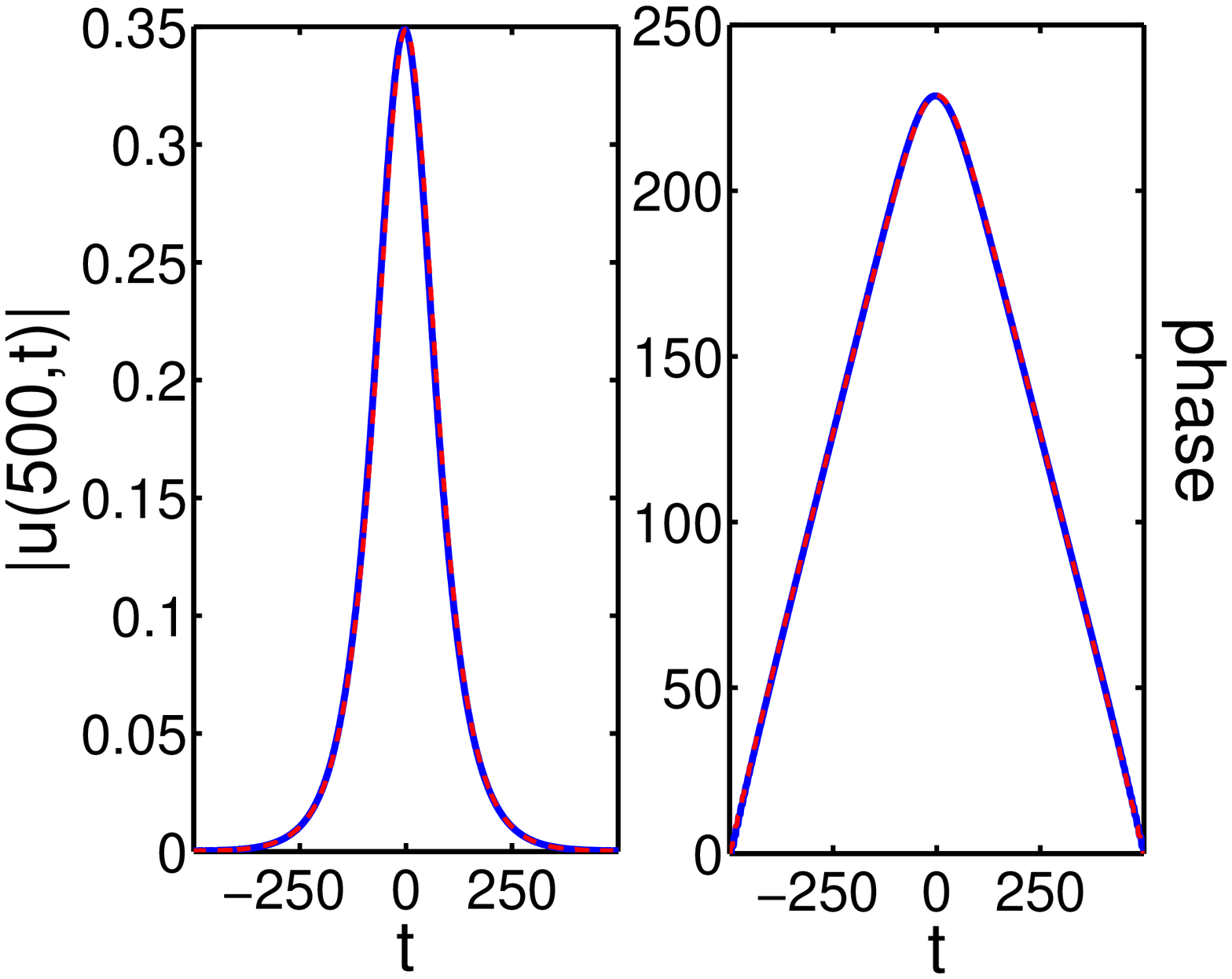}
    \caption{(Color online) Mode-locking in the normal regime (a) under the
    unperturbed PES equation with  $\beta=\gamma=R=0$
    and (b) under the PES with the higher order effects included. The bottom
    figure depicts the final pulse. The dashed line is the soliton of the
    unperturbed PES equation.}
    \label{similariton}
\end{figure}

Recent
experiments indicate that localized modes-- solitons in the normal regime are robust. They
are not affected by
perturbations \cite{oktem} such as TOD and self steepening; this is also in agreement with
the
observation that nonlinear phase shifts (in terms of Raman gain here) can
compensate for TOD or self-steepening  \cite{zhou}. To demonstrate this
phenomena, we
evolve a unit gaussian under Eq. \eqref{pes}, with all parameters as before
($\tau=l=0.1$, $\beta=\gamma=R=0.01$) only now with $g=1.5$. The reason for
this increase in gain is the nature/size of these pulses (cf. Fig.
\ref{similariton}).
Since these pulses are much wider one expects that more energy (gain) is
required for pulses to lock to noticeable amplitudes. Furthermore, in Fig.
\ref{similariton} (bottom) we depict the resulting pulse and its
corresponding phase. The dashed line indicates the soliton of the unperturbed
(i.e. $\beta=\gamma=R=0$) PES equation. Indeed the  experimental observations \cite{ilday} on the shape and phase of the pulses are met. We also confirm subsequent observations
that unlike the solitons of the anomalous regime these wide
localized pulses are resistant to perturbations even when allowed to propagate for greater
distances.

\subsection{Asymptotic theory}

Following the theory developed in Ref. \cite{horikis4} we
write the solution of Eq. (\ref{pes}) in the form $\psi(z,t)=A\exp(i\mu
z+i\theta)$, where $A=A(t)$ and $\theta=\theta(t)$ are the pulse amplitude and
phase respectively. Since these pulses are slowly varying in $t$,
with large phase, we can introduce a slow time scale in the equation
and using perturbation theory, the soliton  system can
be  reduced to simpler ordinary differential equations for the amplitude and the
phase of the pulse. Substituting into the equation and equating real and imaginary
parts we get
\begin{subequations}
\begin{gather}
- \mu A - \frac{{{d_0}}}{2}({A_{tt}} - A\theta _t^2) + {A^3} =  - \frac{\tau }{{1 + \epsilon E}}(2{A_t}{\theta _t} + A{\theta _{tt}})\nonumber \\+ \beta (A{\theta _{tt}} - 3{A_{tt}}{\theta _t} - 3{A_t}{\theta _{tt}} - A{\theta _{ttt}}) + \gamma {A^3}{\theta _t} + 2R{A^2}{A_t}\label{real}\\
- \frac{{{d_0}}}{2}(2{A_t}{\theta _t} + A{\theta _{tt}}) = \frac{g}{{1 + \epsilon E}}A + \frac{\tau }{{1 + \epsilon E}}({A_{tt}} + A\theta _t^2)\nonumber\\ - \frac{l}{{1 + \delta {A^2}}}A+ \beta (3{A_t}\theta _t^2 - 3A{\theta _t}{\theta _{tt}} + {A_{ttt}}) - 3\gamma {A^2}{A_t} \label{imag}
\end{gather}
\label{real.imag}
\end{subequations}
where we have replaced $d_0\rightarrow -d_0<0$ since we are in the normal dispersion regime. We then take  the characteristic time length of the pulse to be such that we
can define a scaling in the independent variable of the form
$\varepsilon=T/\varepsilon$ or $T=\varepsilon t$ so that $R=R(\varepsilon t)$, $\theta_t=O(1)$ (i.e. $\theta$ is large) and
$\theta_{tt}=O(\varepsilon)$ where $\varepsilon \ll 1$. Furthermore in the above simulations $\beta$, $\gamma$ and $R$ are all $O(\varepsilon)$. Then Eq. (\ref{real})
becomes
\[
-\mu A+\frac{d_0}{2}A\theta_t^2+A^3=\frac{d_0}{2}\varepsilon^2
A_{TT}-\frac{\tau}{1+\epsilon E}(2\varepsilon A_T\theta_t+A\theta_{tt})
\]
while the leading order equation is
\begin{gather}
\theta_t^2=\frac{2}{d_0}(\mu-A^2) \label{real2}
\end{gather}
Using the same argument and this newly derived Eq. (\ref{real2})  we then find that
Eq. (\ref{imag}) to leading order reads
\begin{align}
A_t=&-\sgn(t)\frac{\sqrt{2(\mu-A^2)/d_0}}{3A^2-2\mu}\nonumber\\&\times\left(
\frac{g}{1+\epsilon E}-\frac{4\tau(\mu-A^2)/d_0}{1+\epsilon E}
-\frac{l}{1+\delta A^2} \right) \label{imag2}
\end{align}
This is now a nonlocal first order differential equation for $A=A(t)$, since
$E=\int_{-\infty}^{\infty} A^2\;dt$. We also note that  imposing  $\theta_t(t=0)=0 \Rightarrow \mu \approx A^2(0)$.

\begin{figure}[!htbp]
    \centering
    \includegraphics[height=2in]{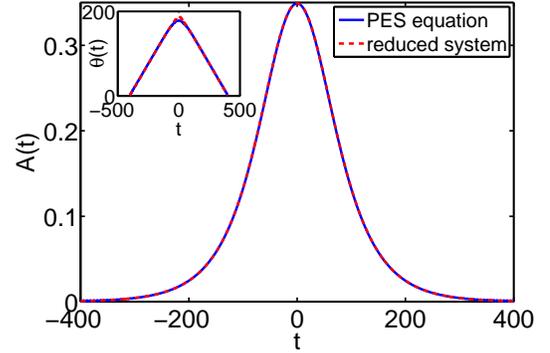}
    \caption{(Color online) Solutions of the complete PES equation and the reduced system.}
    \label{compare}
\end{figure}

To remove the nonlocality another condition is needed and it is based on the
singular points of Eq. (\ref{imag2}). Recall, $A(t)$ is a decaying function in
$t\in[0,+\infty)$ and $\mu \approx A^2(0)\ge A(t)^2$. Thus there exists a point in
$t$ such that the denominator in the equation becomes zero. To remove the
singularity we require that the numerator of the equation is also zero at the same
point which leads to
\[
1+\epsilon E=\frac{1}{l}\left( g-\frac{2\tau}{3d_0}\mu \right)\left(
1+\frac{2\delta}{3}\mu\right)
\]
Thus Eq. (\ref{imag2}) is now a  first order equation which can be solved by
standard numerical methods and analyzed by phase plane methods. The resulting
solutions from the  PES equation  and the reduced equations are compared in Fig.
\ref{compare}.

\subsection{Higher order states}

\begin{figure}[!htbp]
    \centering
    \includegraphics[width=3in]{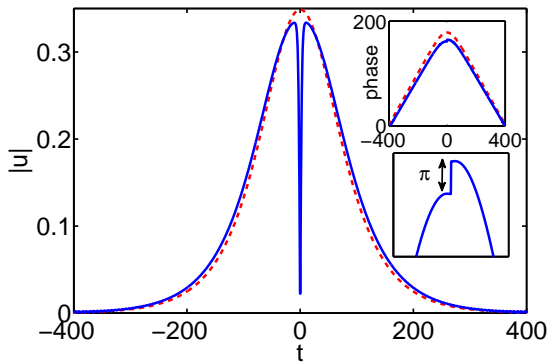}
    \caption{(Color online) A typical bisoliton of the PES. Dashed lines
    correspond to the soliton and phase of the localized pulse of the
    unperturbed PES equation.}
    \label{bisoliton2}
\end{figure}

Remarkably, in the normal regime the PES also exhibits higher-order solutions
in term of bi-solitons \cite{chong}. These are pairs of regular solitons
whose peak amplitudes have a $\pi$ difference in phase and an appropriate
separation. They, also, differ significantly from the higher-order solutions
of the classical NLS and the dispersion managed solitons since they do not
exhibit any oscillatory or breathing behavior as they propagate in the
cavity. A typical such mode is given in Fig. \ref{bisoliton2}. The dashed
line corresponds to the relative ``ground state", i.e. the localized pulse of
the unperturbed PES equation.

\begin{figure}[!htbp]
    \centering
    \includegraphics[width=2.5in]{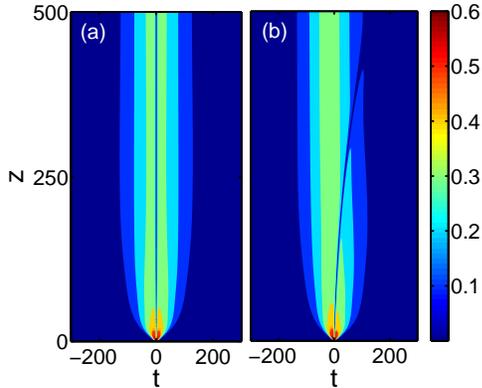}
    \caption{(Color online) Bisoliton evolution (a) under the PES equation;
    (b) under Eq. \eqref{pes}.}
    \label{bisoliton}
\end{figure}

Higher order states and interaction of these structures have been discussed
in Ref. \cite{horikis6}. It is, however, interesting to see their response to
TOD and Raman gain. In order to generate these modes we use an antisymmetric
initial condition such as $u(0,t)=t\exp(-t^2)$. The resulting evolution,
under Eq. \eqref{pes}, is shown in Fig. \ref{bisoliton}.

Under these higher order effects the dip characterizing the bi-soliton, which
separates the two solitons, behaves as a dark soliton on a nearly constant
background and starts  moving until it disappears; see Fig.
\ref{bisoliton}.
This is reminiscent of the dark solitons generated in Bose-Einstein
condensates \cite{dfrantz}. We will further explore this elsewhere.

\section{Conclusions}

To conclude, we have studied the response of a PES model with additional
higher order terms:  third order
dispersion, self-steepening and Raman gain. In the anomalous dispersion
regime, it is found that the PES equation provides the mode-locking mechanism
and after that the  resulting pulses propagate  in a manner similar to
solitons of a
perturbed classical NLS equation.  Perturbative soliton  theory confirms
numerical observations. On the other hand, in the normal dispersion regime,
ground state pulses are more robust and are largely unaffected by these
perturbing influences as
also observed in experiments. However, this is not the case with higher
order modes which are unstable when perturbed.

\acknowledgments
This research was partially supported by the United States Air Force Office
of Scientific Research (USAFOSR) under grant FA9550-12-1-0207.
\endacknowledgments

%\bibliographystyle{osajnl}
%\bibliography{biblio}

\end{document}